\documentstyle[sprocl]{article}

\input{psfig.sty}
\def\Journal#1#2#3#4{{#1} {\bf #2}, #3 (#4)}

\def\apj{\em Astrophys. J.}
\def\apjs{\em Astrophys. J. Suppl.}

\def\MNRAS{\em MNRAS}
\def\NCA{\em Nuovo Cimento}

\def\npps{\em Nucl. Phys. (Proc. Suppl.)}
\def\PLB{{\em Phys. Lett.}  B}
\def\ppnp{\em Prog. in Part. and Nucl. Phys.}
\def\PRL{\em Phys. Rev. Lett.}
\def\PRC{{\em Phys. Rev.} C}
\def\PRD{{\em Phys. Rev.} D}
\def\Science{\em Science}
\def\sjnp{\em Sov. J. Nucl. Phys.}

\def\be{\begin{equation}}
\def\ee{\end{equation}}
\def\bea{\begin{eqnarray}}
\def\eea{\end{eqnarray}}

\def\gtwid{\mathrel{\raise.3ex\hbox{$>$\kern-.75em\lower1ex\hbox{$\sim$}}}}
\def\ltwid{\mathrel{\raise.3ex\hbox{$<$\kern-.75em\lower1ex\hbox{$\sim$}}}}

\begin{document}

\title{NEUTRINO DARK MATTER}

\author{ DAVID O. CALDWELL }

\address{Institute for Nuclear and Particle Astrophysics and Cosmology and\\
Physics Department, University of California,
Santa Barbara,\\ CA 93106-9530, USA}


\maketitle\abstracts{
There is a puzzling contradiction: direct observations favor a low-mass
density universe, but the only model which fits universe structure over
more than three orders of magnitude in distance scale has a mix of hot
(neutrino) and cold dark matter constituting a critical density universe.
If all present indications for neutrino mass are valid, that hot dark
matter is shared by two neutrino species ($\nu_\mu$ and $\nu_\tau$).
These results also require at least one light sterile neutrino to exist
to explain the solar $\nu_e$ deficit ($\nu_e\to\nu_s$), so that
$\nu_\mu\to\nu_\tau$ accounts for the atmospheric neutrino anomaly, with
$\bar\nu_\mu\to\bar\nu_e$ being observed in the LSND experiment.  This
experiment, when analyzed appropriately, does not conflict with any others
and is compatible with the mass difference needed for dark matter.  Support
for this mass pattern is provided by the need for a sterile neutrino to
make possible heavy-element nucleosynthesis in supernovae.}
  
\section{The Dark Matter Conundrum}

Observations of high-redshift Type 1a supernovae, evolution of galaxy clusters,
high baryon content of galaxies, lensing arcs in clusters, dynamical estimates
from infrared galaxy surveys, and especially the existence of galaxies at
very high redshift\cite{ref:1} all indicate that the matter density of the
universe is less than critical (i.e., $\Omega_m<1$).  On the other hand,
the only model which fits the data on the cosmic microwave background
anisotropies and the large-scale distribution of galaxies and clusters is
a model (CHDM) having $\Omega_m=1$, with 70\% cold dark matter, 20\% hot
dark matter, and 10\% baryonic matter.\cite{ref:2}  These structure data
covering three orders of magnitude in distance scale seemingly exclude
models of an open universe (low $\Omega_m$) or one adding a cosmological
constant ($\Lambda$) to provide critical energy density
($\Omega_m+\Omega_\Lambda=1$), since the fit probabilities are
$\sim10^{-4}$--$10^{-5}$.  Extending the fits another order of magnitude
into the smaller scale non-linear regime makes the discrepancy between
CHDM and the low $\Omega_m$ models even greater.\cite{ref:3}  Another
measure of structure, the probability of voids, also strongly favors CHDM
over the low $\Omega_m$ models.\cite{ref:4}  Adding some neutrinos to the
latter helps only a little.\cite{ref:5}

This clearly is an important conflict which may get resolved by better
observations or by the discovery of some new factor which reconciles the
observations.  At this time for help in knowing whether neutrion dark
matter exists, we must turn to observations of neutrino mass.

\section{Evidence for Neutrino Mass\label{sec:2}}

Three types of observation give evidence for neutrino mass.  Two of these
are of importance here only in establishing a likely pattern of neutrino
masses, and hence these will be treated cursorily, but the third is of
direct relevance for hot dark matter and will therefore be discussed more
thoroughly.

The first is the deficit of electron neutrinos from the sun observed by
four experiments.  The four are of three types, covering different $\nu_e$
energy ranges, and hence sampling different contributions from various
nuclear processes producing neutrinos.  These show an energy-dependent
discrepancy exemplified by an apparent lack (the best fit being a negative
flux) of neutrinos from $^7$Be and a finite flux from $^8$B neutrinos, yet
$^8$B is produced by $^7\rm Be+p\to\/^8B+\gamma$.  This problem cannot be
avoided by one of the experiments being wrong.  A good solution is provided
if the $\nu_e$ oscillates into $\nu_\mu$, $\nu_\tau$, or $\nu_s$, a sterile
neutrino.  While this can be a vacuum oscillation, requiring a mass-squared
difference $\Delta m^2_{ei}\sim10^{-10}$ eV$^2$ and large mixing between
$\nu_e$ and the other neutrino, more likely is a matter-enhanced
MSW\cite{ref:6} type of oscillation.  For a $\nu_\mu$ or $\nu_\tau$ final
state, $\Delta m^2_{ei}\sim10^{-5}$ eV$^2$ and mixings either
$\sin^22\theta_{ei}\sim6\times10^{-3}$ or $\sim0.6$, while only the former
is allowed for $\nu_s$.

The second observation explainable by oscillations and hence providing
evidence for neutrino mass has now been furnished by three water Cherenkov
detectors and two tracking calorimeters.  The ratio of $\nu_\mu$ to
$\nu_e$ produced in the atmosphere is found to be roughly half that expected.
These underground experiments observe the $\mu$ and $e$ products and
utilize $R=(\mu/e)_{\rm Data}/(\mu/e)_{\rm Calc.}$ to reduce flux
uncertainties.  While the statistical evidence for $R$ being less than unity
is now quite compelling, it is the difference in the angular distribution
of the $\mu$ and $e$ events which provides the primary evidence that the
explanation is neutrino oscillations, giving
$\Delta m^2_{\mu i}\sim2\times10^{-3}$ eV$^2$ and $\sin^22\theta_{\mu i}\sim1$.
These angular distributions, as well as lack of $\nu_e$ disappearance observed
in a nuclear reactor experiment,\cite{ref:7} rule out $\nu_\mu\to\nu_e$ as
the predominant process.  Quite unlikely is a $\nu_\mu\to\nu_s$ explanation,
since the large mixing angle would cause the sterile neutrino to be brought
into equilibrium in the early universe, providing a fourth neutrino which
would alter the expansion rate in the era of nucleosynthesis, spoiling the
agreement between calculations and the observed abundances of light elements.
There is now fair concordance between $^4$He abundance values\cite{ref:8}
and the primordial D/H ratio,\cite{ref:9} reinstating the three-neutrino
limit.  A possible way\cite{ref:10} around this appears to be ruled
out.\cite{ref:11}  Thus the only viable explanation has $\nu_\mu\to\nu_\tau$
as the dominant process.

The third evidence for neutrino mass is from the LSND accelerator experiment,
and the mass difference observed is directly relevant for the issue of
neutrino dark matter.  This experiment used a decay-in-flight $\nu_\mu$ beam
of up to $\sim180$ MeV from $\pi^+\to\mu^+\nu_\mu$ and a decay-at-rest
$\bar\nu_\mu$ beam of less than 53 MeV from the subsequent
$\mu^+\to e^+\nu_e\bar\nu_\mu$.  The 1993+1994+1995 data sets\cite{ref:12}
included 22 events of the type $\bar\nu_ep\to e^+n$, expected from
$\bar\nu_\mu\to\bar\nu_e$, which was based on identifying an electron using
Cherenkov and scintillation light that was tightly correlated with a $\gamma$
($<0.6$\% accidental rate) from $np\to d\gamma$ (2.2 MeV).  Only $4.6\pm0.6$
such events were expected from backgrounds.  The chance that these
data, using a water target, result from a fluctuation is $4\times10^{-8}$. 
Note especially that these data were restricted to the energy range 36 to 60
MeV to stay below the $\bar\nu_\mu$ endpoint and to stay above the region where
backgrounds are high due to the $\nu_e\/^{12}{\rm C}\to e^-X$ reaction.  In
plotting $\Delta m^2$ vs.\ $\sin^22\theta$, however, events down to 20 MeV
were used to increase the range of $E/L$, the ratio of the neutrino's energy
to its distance from the target to detection.  This was done because the plot
employed was intended to show the favored regions of $\Delta m^2$, and all
information about each event was used.  A likelihood analysis was utilized
with contours which would be 90\% and 99\% likelihood levels, if this were a
Gaussian likelihood distribution, which it is not because its integral is
infinite.  Those contours have been widely misinterpreted as confidence
levels---which they certainly are not---because they were plotted along with
confidence-level limits from other experiments.  The comparison of LSND
``favored regions'' contours with confidence-level limits, especially
because of the use of the 20--36 MeV region for the LSND data, has led to
totally incorrect conclusions.

\begin{figure}
\parbox[t]{5.25cm}
{\psfig{figure=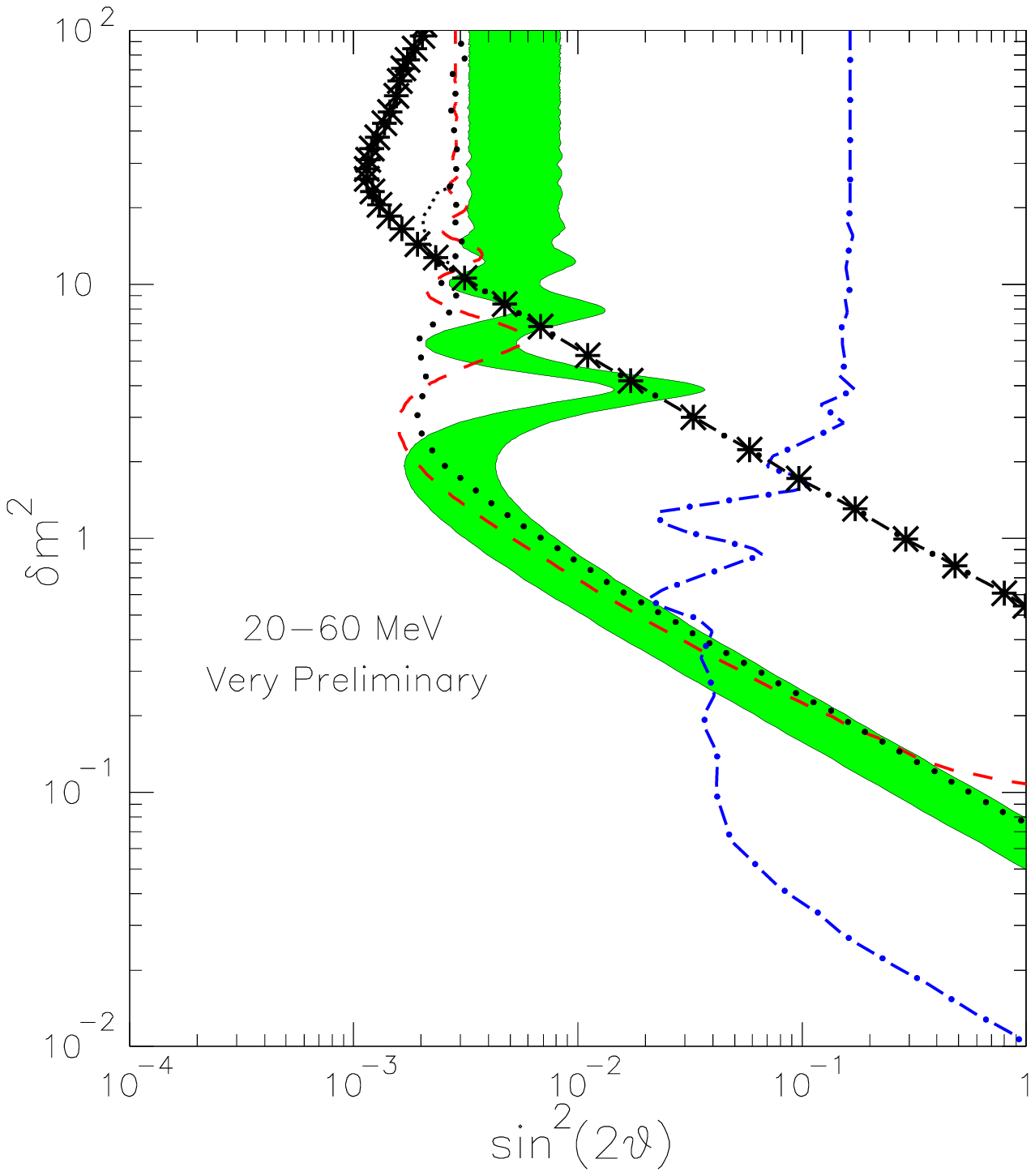,height=6cm}
\caption{LSND upper and lower 90\% Bayesian confidence limits for
$\bar\nu_\mu\rightarrow \bar\nu_e$ oscillations using 1993-1997 data with
$20<E_e<60$ MeV.  Also shown is the 90\% Bayesian confidence upper limit
from KARMEN as of summer, 1998 (dashed), as well as upper limits from NOMAD
(dash-dot-star), E776 (dotted), and Bugey (dash-dot).
\label{fig:1}}
}
\hspace{-5mm}
\parbox[t]{5.25cm}
{\psfig{figure=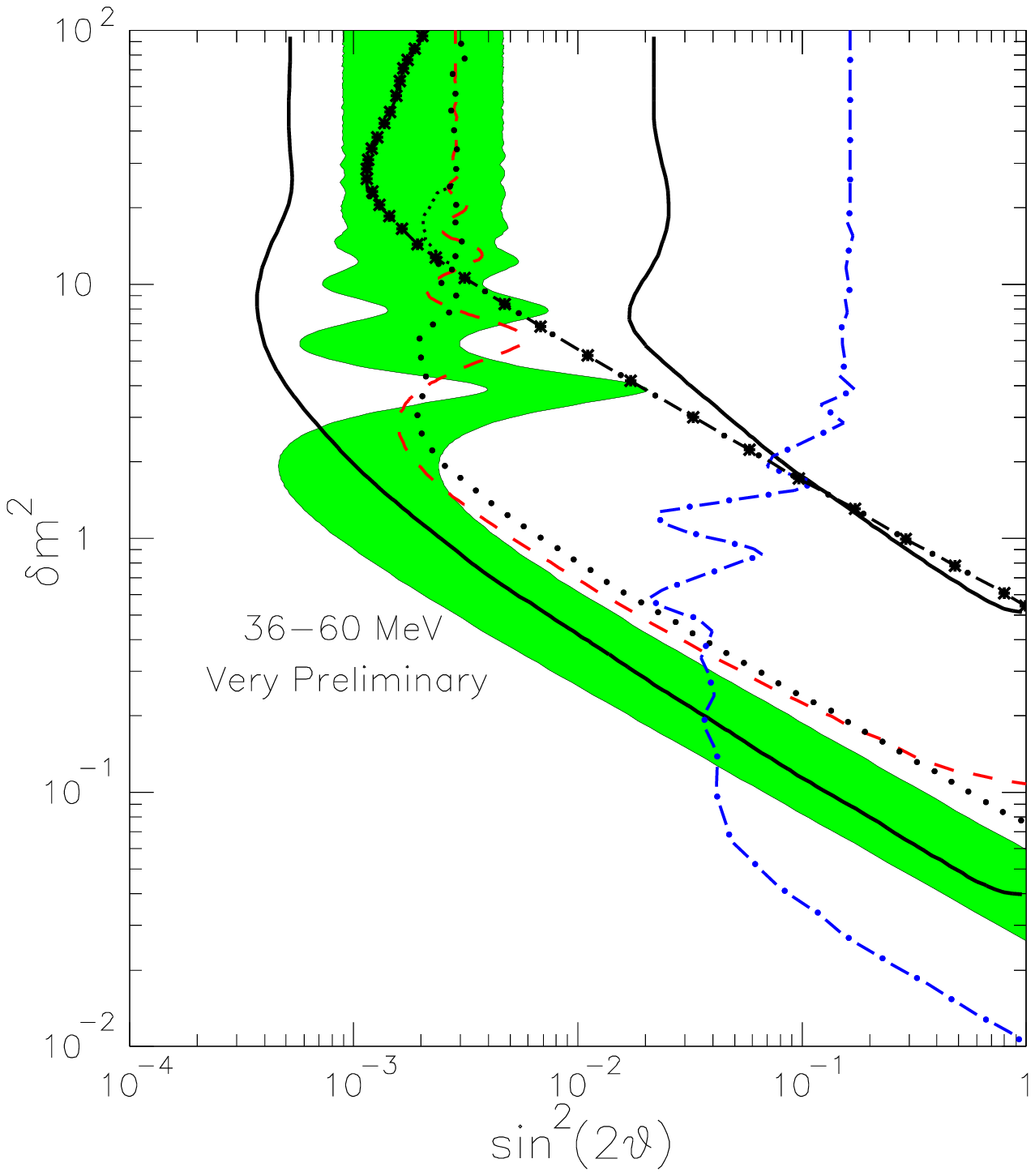,height=6cm}
\caption{Same as Figure \ref{fig:1}, but from $36<E_e<60$ MeV data.
The added curves show the 80\% confidence band for the LSND $\nu_\mu\to\nu_e$
result.
\label{fig:2}}
}
\end{figure}

In particular, in the summer of 1998 when the KARMEN\cite{ref:13} experiment
with improved shielding was observing no events, such a comparison was often
said to rule out the LSND result.  While KARMEN is reported now to have
observed three events, making their results less exclusive, we use here the
zero-event results of last summer to illustrate the difference made by a
fairer comparison of the experiments analyzed by the same (Bayesian) method.
To do this it is necessary not to use the full E/L information, so this is
not the way to determine favored regions of $\Delta m^2$.  Figure~\ref{fig:1}
still shows large regions of conflict if the LSND data from 1993 through 1997
are used for the full 20--60 MeV range of electron energies.  On the other
hand, if the region originally used to establish an effect (36--60 MeV) is
used, exluding the higher background 20--36 MeV region, then Fig.~\ref{fig:2}
shows no conflict over a very large $\Delta m^2$ range.  The 1996 through
1998 data adds 50\% more beam-on events to the 36--60 MeV data but makes the
total beam-off background 2.5 times worse because data-taking was parasitic
at a slow rate using an iron target.  Especially if the 20--36 MeV region is
used, the decreased signal/background ratio in the newer data distorts the
energy spectrum, making the higher $\Delta m^2$ values desirable for dark
matter erroneously appear less likely.

Also shown in Fig.~\ref{fig:2} is the LSND $\nu_\mu\to\nu_e$
result\cite{ref:14} which although quite broad tends to favor these higher
$\Delta m^2$ values.  This broadness results from the greater background in
this case, primarily because the observed process ($\nu_eC\to e^-X$) gives
only one signal instead of the two available in the $\bar\nu_\mu\to\bar\nu_e$
case.  While the fluctuation probability for $\nu_\mu\to\nu_e$ is only
$\sim10^{-2}$, the two ways of detecting oscillations are essentially
independent, providing some confirmation that a real effect is being observed.

\section{The Need for Two-Neutrino Dark Matter and a Sterile Neutrino}

If, as argued above, the atmospheric $\nu_\mu/\nu_e$ ratio is explained by
$\nu_\mu\to\nu_\tau$, one-neutrino dark matter is ruled out, since
$\Delta m^2_{\mu\tau}\sim10^{-3}$ eV$^2$.  Note that the total neutrino mass
needed is 94 $\Omega_\nu h^2\sim5$ eV$^2$ for 20\% neutrinos and $\Omega_m=1$
with $h=0.5$, or $\Omega_m=0.6$ with $h=0.65$, where $h$ is the Hubble constant
in units of 100 km$\cdot$s$^{-1}\cdot$Mpc$^{-1}$.  If the LSND result were
invalid, three-neutrino dark matter would be possible, with $\nu_e$, $\nu_\mu$,
and $\nu_\tau$ nearly degenerate in mass.\cite{ref:15}  As shown above,
however, there is no conflict between the LSND result and other experiments
for the mass range desired for hot dark matter.

That leaves two-neutrino dark matter.  This scheme\cite{ref:15} requires
four neutrinos, with the solar deficit explained by $\nu_e\to\nu_s$ (and
both neutrinos quite light) the atmospheric effect due to $\nu_\mu\to\nu_\tau$
(both of which are heavier and share the dark matter role) and the LSND
$\nu_\mu\to\nu_e$ demonstrating the mass difference between these two nearly
mass-degenerate doublets.  Note that the solar $\nu_e\to\nu_s$ is for the
small mixing angle (or vacuum oscillation), so $\nu_s$ does not affect
nucleosynthesis.  The original motivation for this mass pattern\cite{ref:15}
preceded LSND and was simply to provide some
hot dark matter, given the solar and atmospheric phenomena.  If LSND is
correct, it becomes the unique pattern.  However, just the $\nu_\mu\to\nu_\tau$
explanation of the atmospheric result alone forces two-neutrino dark matter.

This neutrino scheme was the basis for simulations\cite{ref:16} which showed
that two-neutrino dark matter fits observations better than the one-neutrino
variety.  The latter produces several problems at a distance scale of the
order of $10h^{-1}$ Mpc, particularly overproducing clusters of galaxies.
Whether the $\sim5$ eV of neutrino mass is in the form of one neutrino
species or two makes no difference at very large or very small scales, but
at $\sim10h^{-1}$ Mpc the larger free streaming length of $\sim5/2$ eV
neutrinos washes out density fluctuations and hence lowers the abundance of
galactic clusters.  In every aspect of simulations done subsequently the
two-neutrino dark matter has given the best results.  For example, a single
neutrino species, as well as low $\Omega_m$ models, overproduce void regions
between galaxies, whereas the two-neutrino model agrees well with
observations.\cite{ref:4}

Two-neutrino dark matter requires at least one other light neutrino which
must not have the normal weak interaction because of the measured width of
the $Z^0$ boson.  Independent information favoring such a sterile neutrino
comes from the excellent neutrino laboratory, the supernova.

While $\Delta m^2_{e\mu}\sim6$ eV$^2$ is desirable for two-neutrino
dark matter, it apparently would cause a conflict with the
production of heavy elements in supernovae.  This $r$-process of rapid neutron
capture occurs in the outer neutrino-heated ejecta of Type II supernovae.  The
existence of this process would seem to place a limit on the mixing of
$\nu_\mu$ and $\nu_e$ because energetic $\nu_\mu\ (\langle E\rangle\approx25$
MeV) coming from deep in the supernova core could convert via an MSW transition
to $\nu_e$ inside the region of the $r$-process, producing $\nu_e$ of much
higher energy than the thermal $\nu_e\ (\langle E\rangle\approx11$ MeV).  The
latter, because of their charged-current interactions, emerge from farther out
in the supernova where it is cooler.  Since the cross section for $\nu_en\to
e^-p$ rises as the square of the energy, these converted energetic $\nu_e$
would deplete neutrons, stopping the $r$-process. 
Calculations\cite{ref:17} of this effect limit $\sin^22\theta$ for
$\nu_\mu\to\nu_e$ to $\ltwid10^{-4}$ for $\Delta m^2_{e\mu}\gtwid2$ eV$^2$, in
conflict with compatibility between the LSND result and a neutrino component of
dark matter.

The sterile neutrino not only solves this problem, but also rescues
the $r$-process itself.  While simulations have found the $r$-process
region to be insufficiently neutron rich,\cite{ref:18} recent
realization of the full
effect of $\alpha$-particle formation has created a disaster for the
$r$-process.\cite{ref:19} The initial difficulty of too low entropy (i.e., too
few neutrons per seed nucleus, like iron) has now been drastically exacerbated
by calculations\cite{ref:19} of the sequence in which all available
protons swallow up neutrons to form $\alpha$ particles, following which
$\nu_en\to e^-p$ reactions create more protons, creating more $\alpha$
particles, and so on.  The depletion of neutrons by making $\alpha$ particles
and by $\nu_en\to e^-p$ rapidly shuts off the $r$-process, and essentially no
nuclei above $A=95$ are produced.

The sterile neutrino would produce two effects.\cite{ref:20}  First, there is
a zone, outside the neutrinosphere (where neutrinos can readily escape) but
inside the $\nu_\mu\to\nu_e$ MSW (``LSND") region, where the $\nu_\mu$
interaction potential goes to zero, so a $\nu_\mu\to\nu_s$ transition can occur
nearby, depleting the dangerous high-energy $\nu_\mu$ population.  Second,
because of this $\nu_\mu$ reduction, the dominant process in the MSW region
reverses, becoming $\nu_e\to\nu_\mu$, dropping the $\nu_e$ flux going into the
$r$-process region, hence reducing $\nu_en\to e^-p$ reactions and allowing the
region to be sufficiently neutron rich.

This description is simplified, since the atmospheric results show that the
$\nu_\mu$ and $\nu_\tau$ mix with a large angle, so wherever ``$\nu_\mu$''
is mentioned, this can equally well be ``$\nu_\tau$''.  In fact, if the
mixing is maximal and the $\nu_\mu$ and $\nu_\tau$ mix equally with the
$\nu_e$, one can show\cite{ref:20} that the $\nu_e$ flux above the second
resonance vanishes totally.  To keep the resonances separate and in the
proper order, they must occur below the weak freeze out radius, where the
weak interactions go out of equilibrium.  This requires a sufficiently large
$\Delta m^2_{e\mu(\tau)}$, and a value like 6 eV$^2$ satisfies this
requirement, enhancing the argument for hot dark matter.

The near concordance for nucleosynthesis between $^4$He abundance and the
D/H ratio alluded to in Section \ref{sec:2} requires that the sterile neutrino
not have much effect on $^4$He abundance.  The $\nu_s$ needed for solar $\nu_e$
depletion does not come into equilibrium in that era, so it has little effect
on the expansion rate, but it could cause a $\nu_e/\bar\nu_e$
asymmetry,\cite{ref:10} changing the $n/p$ ratio and hence altering the
amount of $^4$He.  A calculation\cite{ref:21} of this effect showed that for
this particular model the change in $^4$He was small, possibly being in the
right range to correct a small remaining discrepancy, if there really is one.

\section{Conclusions}

A neutrino component of dark matter appears very probable, both from the
astrophysics and particle physics standpoints.  Despite the evidence for
$\Omega_m<1$, the one model which fits universe structure has $\Omega_m=1$,
with 20\% neutrinos and most of the rest as cold dark matter.  Open universe
and low-density models with a cosmological constant give extremely bad fits.
This conflict should be the source of future progress, but since there are
$10^2/\rm cm^3$ of neutrinos of each active species left over from the early
universe, the ultimate answer on neutrino dark matter will come from
determinations of neutrino mass.  While the solar and atmospheric evidences
for neutrino mass are important, the crucial issue is the much larger
mass-squared difference observed by the LSND experiment.  In the mass region
needed for dark matter, no other experiment excludes the LSND result, if
data from the different experiments are compared using the same procedures.

The resulting mass pattern, $\nu_e\to\nu_s$ for solar, $\nu_\mu\to\nu_\tau$
for atmospheric, and $\nu_\mu\to\nu_e$ for LSND, requires a sterile neutrino
and provides two-neutrino ($\nu_\mu$ and $\nu_\tau$) dark matter.  This
form of dark matter fits observational data better than the one-neutrino
variety.  Furthermore, the sterile neutrino appears to be necessary to
rescue the production of heavy elements by supernovae.  This particular mass
pattern does not cause any difficulty with the present near concordance in
primordial light element abundances, and it could even help if there is a
small discrepancy.  In short, this four-neutrino pattern agrees with all
current neutrino mass information and hence makes more likely the existence
of hot dark matter.

\section*{Acknowledgments}
I wish to thank Steven Yellin for producing the figures.  This work was
supported in part by the U.S.~Department of Energy.


\section*{References}
\begin{thebibliography}{99}
\bibitem{ref:1}R.S. Somerville, J.R. Primack, and S.M. Faber,
astro-ph/9806228; MNRAS in press.
\bibitem{ref:2} E. Gawiser and J. Silk, \Journal{\Science}{280}{1405}{1988};
M.A.K. Gross, R.S. Somerville, J.R. Primack, J. Holtzman, and A.A. Klypin,
\Journal{\MNRAS}{301}{81}{1998}.
\bibitem{ref:3} J.R. Primack and A. Klypin, \Journal{\npps}{B51}{30}{1996}.
\bibitem{ref:4} S. Ghigna {\it et al.}, \Journal{\apj}{479}{580}{1997};
{\it ibid.} {\bf 437}, L71 (1994).
\bibitem{ref:5} J.R. Primack and M.A.K. Gross, astro-ph/9810204v4 (1998).
\bibitem{ref:6} L. Wolfenstein, \Journal{\PRD}{17}{2369}{1978};
\Journal{\PRD}{20}{2634}{1979}; S.P. Mikheyev and A. Yu. Smirnov,
\Journal{\sjnp}{42}{913}{1985}; \Journal{\NCA}{9C}{17}{1986}.
\bibitem{ref:7} M. Apollonio {\it et al.}, \Journal{\PLB}{420}{397}{1998}.
\bibitem{ref:8} Y.I. Izotov and T.X. Thuan, \Journal{\apj}{500}{188}{1998}.
\bibitem{ref:9} S. Burles and D. Tytler, \Journal{\apj}{499}{699}{1998};
S.A. Levshakov, W.H. Kegel, and F. Takahara, \Journal{\apj}{499}{L1}{1998}.
\bibitem{ref:10} R. Foot and R.R. Volkas, \Journal{\PRD}{56}{6653}{1997};
{\bf D55}, 5147 (1997); R. Foot, hep-ph/9809315 (1998); R. Foot and
R.R. Volkas, astro-ph/9811067 (1998).
\bibitem{ref:11} X. Shi and G.M. Fuller, astro-ph/9810075 (1998);
astro-ph/9812232 (1998).
\bibitem{ref:12} C. Athanassopoulos {\it et al.},
 \Journal{\PRC}{54}{2685}{1996}; \Journal{\PRL}{77}{3082}{1996}.
\bibitem{ref:13} G. Drexlin {\it et al.}, \Journal{\ppnp}{32}{375}{1994};
B. Armbruster {\it et al.}, \Journal{\npps}{B38}{235}{1995}; B. Zeitnitz at
Neutrino `98.
\bibitem{ref:14} C. Athanassopoulos {\it et al.},
 \Journal{\PRL}{81}{1774}{1998}.
\bibitem{ref:15} The three- and four-neutrino mass schemes to give dark matter
were first proposed in D.O. Caldwell, {\it Perspectives in Neutrinos, Atomic
Physics and Gravitation}, Editions Fronti\`eres, Gif-sur-Yvette, France,
1993, p. 187; D.O. Caldwell and R.N. Mohapatra, \Journal{\PRD}{48}{3259}{1993}.
\bibitem{ref:16} J.R. Primack, J. Holtzman, A. Klypin, and D.O. Caldwell,
\Journal{\PRL}{74}{2160}{1995}.
\bibitem{ref:17} Y.-Z. Qian {\it et al.}, \Journal{\PRL}{71}{1965}{1993};
Y.-Z. Qian and G.M. Fuller, \Journal{\PRD}{51}{1479}{1995}; G. Sigl,
\Journal{\PRD}{51}{4035}{1995}.
\bibitem{ref:18} R.D. Hoffman, S.E. Woosley, and Y.-Z. Qian,
\Journal{\apj}{482}{951}{1996}; B.S. Meyer and J.S. Brown,
\Journal{\apjs}{112}{199}{1997}.
\bibitem{ref:19} B.S. Meyer, G.C. McLaughlin, and G.M. Fuller,
{\em Phys. Rev. C}, in press (1998).
\bibitem{ref:20} D.O. Caldwell, G.M. Fuller, and Y.-Z. Qian (to be published).
\bibitem{ref:21} N.F. Bell, R. Foot, and R.R. Volkas,
 \Journal{\PRD}{58}{105010}{1998}.

\end{thebibliography}
\end{document}